\documentclass[runningheads]{llncs}

\usepackage[T1]{fontenc}
\def\doi#1{\href{https://doi.org/\detokenize{#1}}{\url{https://doi.org/\detokenize{#1}}}}

\usepackage{multirow}
\usepackage{tabularx}
\usepackage{amssymb}
\usepackage{amsmath}
\usepackage{bm}
\usepackage{graphicx}
\usepackage{amsfonts}
\usepackage{booktabs}
\usepackage{array}
\usepackage{algorithm} 
\usepackage[algo2e]{algorithm2e} 
\usepackage{algpseudocode}
\usepackage{siunitx}
\usepackage{cite}
\usepackage{hyperref}
\usepackage{mathtools}
\usepackage[misc]{ifsym}
\usepackage{pifont}
\def\Plus{\texttt{+}}
\usepackage{gensymb}
\newcolumntype{C}[1]{>{\centering\let\newline\\\arraybackslash\hspace{0pt}}m{#1}}
\usepackage{listings}
\begin{document}
\title{SegHeD: Segmentation of Heterogeneous Data for Multiple Sclerosis Lesions with Anatomical Constraints}
\titlerunning{SegHeD}
%
\author{Berke Doga Basaran\inst{1,2}\textsuperscript{(\Letter)} \and 
Xinru Zhang\inst{3,4} \and
Paul M. Matthews\inst{3,5,6} \and \\
Wenjia Bai\inst{1,2,3}}
%

\authorrunning{B.D. Basaran et al.}
%
\institute{Department of Computing, Imperial College London, London, UK \and
Data Science Institute, Imperial College London, London, UK \\
\email{\Letter \hspace{.1cm} bdb19@imperial.ac.uk} \and
Department of Brain Sciences, Imperial College London, London, UK \and 
School of Integrated Circuits and Electronics, Beijing Institute of Technology, Beijing, CN \and
UK Dementia Research Institute, Imperial College London, London, UK \and
Rosalind Franklin Institute, Didcot, UK}
\maketitle
\begin{abstract} 
Assessment of lesions and their longitudinal progression from brain magnetic resonance (MR) images plays a crucial role in diagnosing and monitoring multiple sclerosis (MS). Machine learning models have demonstrated a great potential for automated MS lesion segmentation. Training such models typically requires large-scale high-quality datasets that are consistently annotated. However, MS imaging datasets are often small, segregated across multiple sites, with different formats (cross-sectional or longitudinal), and diverse annotation styles. This poses a significant challenge to train a unified MS lesion segmentation model. To tackle this challenge, we present SegHeD, a novel multi-dataset multi-task segmentation model that can incorporate heterogeneous data as input and perform all-lesion, new-lesion, as well as vanishing-lesion segmentation. Furthermore, we account for domain knowledge about MS lesions, incorporating longitudinal, spatial, and volumetric constraints into the segmentation model. SegHeD is assessed on five MS datasets and achieves a high performance in all, new, and vanishing-lesion segmentation, outperforming several state-of-the-art methods in this field.
\end{abstract}

\section{Introduction} 
Multiple sclerosis (MS) is an inflammatory and demyelinating neurological disorder affecting the central nervous system. Segmenting and quantifying MS lesions from brain MR images plays a crucial role in clinical diagnosis and research. While many lesion segmentation methods, such as those based on machine learning, have been proposed in recent years, they are often trained on well-curated datasets with a consistent format~\cite{Carass2015, Commowick2018, Commowick2021, Eisenmann2022}. Clinical MS studies involve collection of heterogeneous imaging data from multiple sites that come in diverse data and annotation formats, which are incompatible for training off-the-shelf segmentation models~\cite{Isensee2021, Hatamizadeh2022}. There is a scarcity of approaches tailored to address lesion segmentation in the presence of heterogeneous data and annotation formats. Efforts to bridge this gap are crucial for advancing the accuracy and applicability of lesion segmentation models in MS research and clinical practice.
In this work, we propose SegHeD, a novel multi-dataset multi-task brain lesion segmentation model that utilises heterogeneous data for training. SegHeD allows learning from both cross-sectional (a single timepoint) and longitudinal (multiple timepoints) images, and learning from diverse annotations for all lesions, new lesions, and less explored vanishing lesions~\cite{prineas199}. We incorporate temporal and volume consistency, and anatomical plausibility for the segmented lesions. Experiments show competitive results against state-of-the-art methods.

\subsection{Related Works}
\subsubsection{Longitudinal lesion segmentation} 
There have been numerous contributions to machine learning methods for cross-sectional brain lesion segmentation~\cite{kamnitsas2017, Zeng2020, basaran2023}. Longitudinal lesion segmentation, which takes advantage of the temporal information within longitudinal imaging data, is relatively less explored. Elliott et al. utilises the difference map between the two timepoints to detect new lesions in the second timepoint~\cite{Elliott2013}. Jain et al. also utilises the difference map and formulates an expectation-maximisation framework to perform joint segmentation for both timepoints~\cite{Jain2016}. Denner et al. incorporates a displacement field to learn spatio-temporal changes between timepoints~\cite{Denner2021}. Basaran et al. employs an nnU-Net~\cite{Isensee2021} along with lesion-aware data augmentation to detect new lesions at the second timepoint~\cite{Basaran2022}.

\subsubsection{Learning from heterogeneous data}
Publicly available datasets for brain lesion segmentation vary in format and annotation protocols. Some are cross-sectional with a single timepoint scan, where others are longitudinal with two or more timepoints. It creates a formidable challenge in developing a universal machine learning model that is inclusive of different formats of data. Wu et al. proposes to learn from heterogenous data for all and new-lesion segmentation by imposing relation regularisation onto all-lesion prediction~\cite{Wu2023}. Liu et al. trains a multi-organ segmentation model using 13 different organ datasets and incorporating CLIP-inspired label encoding~\cite{Liu2023}. Shi et al. proposes a marginal loss and an exclusion loss to learn a single multi-organ segmentation model from partially annotated datasets~\cite{Shi2021}.

\subsubsection{Enforcing anatomical plausibility}
The incorporation of anatomical plausibility within machine learning models enhances the reliability and trustworthiness of models for clinical use. Dalca et al. proposes a model with a variational auto-encoder to learn location-specific priors for brain structure segmentation~\cite{Dalca2018}. Strumia et al. constrains the lesion segmentation to be within the white matter using a geometric brain model~\cite{Strumia2016}. Finally, Hirsh et al. employs a multi-prior network with tissue probability maps for MRI head anatomy segmentation~\cite{Hirsch2021}.

\subsection{Contributions} 
The contributions of the proposed method, SegHeD, are three-fold: (1) It is a general framework that accounts for heterogeneous data of different data formats and annotation protocols. (2) It simultaneously performs multiple tasks, including vanishing-lesion segmentation, a task rarely accounted for in previous research. (3) It incorporates domain knowledge about MS lesion segmentation, including temporal consistency and anatomical plausibility.

\section{Method}
\subsection{Overall architecture}
\begin{figure}[h!] \centering
\includegraphics[width=.95\textwidth]{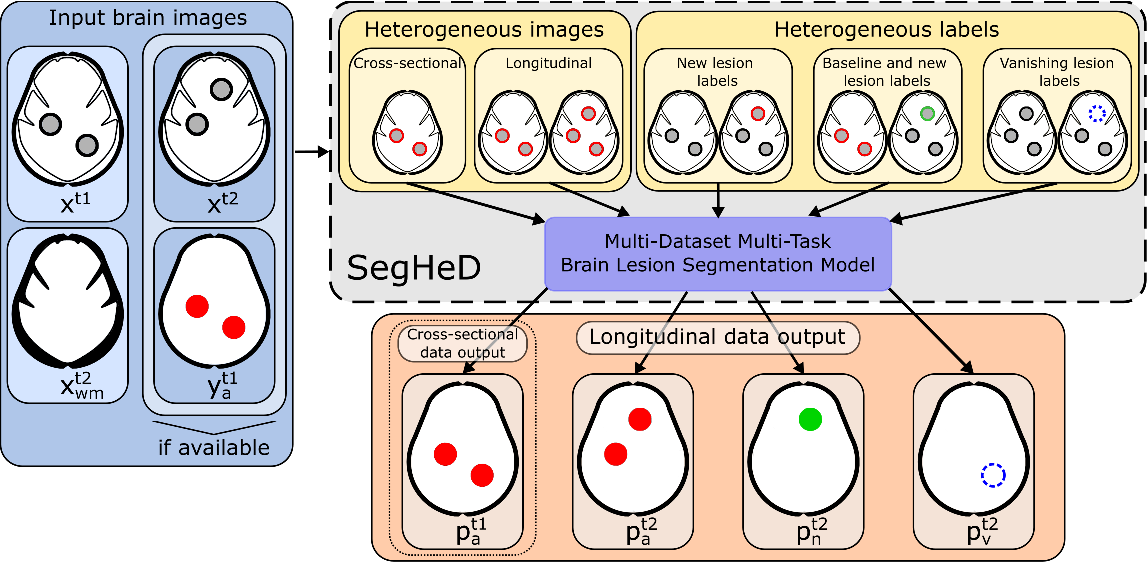}
\caption{Visualisation of the proposed framework. SegHeD learns from heterogeneous datasets varying in image and label formats. It can analyse cross-sectional and longitudinal data, and segment all (red), new (green), and vanishing (dashed blue) lesions.} 
\label{fig:framework}
\end{figure}

SegHeD is a universal model that can learn from heterogeneous MS imaging data. In terms of data heterogeneity, we account for two data formats: cross-sectional and longitudinal. In terms of label heterogeneity, we account for three annotations protocols: all-lesion annotation, new-lesion annotation which only annotates new lesions in the second timepoint scan, and vanishing-lesion annotation which only annotates lesions that vanish in the second timepoint. The latter two protocols focus on the longitudinal evolution of lesions and avoid annotating all lesions to save time in practice. SegHeD aims to learn and perform all three annotation tasks. Figure~\ref{fig:framework} illustrates the overall framework of the method. 

We assume that the imaging data has two timepoints. SegHeD takes 4 input channels: a first timepoint scan (baseline), $x^{t1}$, a second timepoint scan (follow-up), $x^{t2}$, an all-lesion label map for the first timepoint scan, $y^{t1}_{a}$, and a white matter mask for the second timepoint scan, $x^{t2}_{wm}$. For heterogeneous datasets, not all the above inputs are available. Where a second timepoint is not available, the first timepoint is passed as the second, $x^{t2} = x^{t1}$. Where a first timepoint all-lesion label map is not available, $y^{t1}_{a}$ is replaced with a zero matrix of the same size as $x^{t1}$. The second timepoint white matter mask, $x_{wm}^{t2}$, can be obtained using the pretrained SynthSeg brain parcellation tool, which is robust in white matter segmentation even when lesions exist~\cite{Billot2023}. SegHeD learns to generate 4 outputs: first timepoint all-lesion segmentation, $p_{a}^{t1}$, second timepoint all-lesion segmentation, $p_{a}^{t2}$, second timepoint new-lesion segmentation, $p_{n}^{t2}$, and second timepoint vanishing-lesions, $p_{v}^{t2}$. It is formulated as Eq.~\eqref{eq:form},
\begin{equation}
\label{eq:form}
    p_{a}^{t1}, p_{a}^{t2}, p_{n}^{t2}, p_{v}^{t2} =  F(x^{t1}, x^{t2}, y_{a}^{t1}, x_{wm}^{t2}),
\end{equation}
where $F$ denotes the SegHeD model. The all-lesion map, $y^{t1}_{a}$, is only input for second timepoint lesion predictions to provide temporal context, and not used for predicting $p_{a}^{t1}$. The model allows for a maximum of two timepoints as inputs due to computational limitations. For datasets where a subject has more than two timepoint scans~\cite{Carass2015}, we implement a sliding window approach across the multiple timepoints to perform training and inference. We implement SegHeD with a 3D V-Net architecture~\cite{Milletari2016}, with a composite training loss function which includes the Dice loss, longitudinal, spatial, and volumetric constraint losses.

\subsection{Anatomical Constraints}
SegHeD is trained using a novel combination of losses we term \textit{anatomical constraints}. We take inspiration from how radiologists analyse longitudinal images for identifying lesions~\cite{Homssi2023}. Anatomical constraints formulate the longitudinal, spatial, and volumetric relations for lesion segmentations at the two timepoints. 

\subsubsection{Longitudinal constraints}
We take advantage of the prior knowledge encoded within the lesion annotation protocol: (1) Predicted new-lesions at the second timepoint, $p_{n}^{t2}$, can not be present in the all-lesion label at the first timepoint, $y_{a}^{t1}$, but must be present in the second timepoint all-lesion label, $y_{a}^{t2}$. (2) Predicted vanishing-lesions at the second timepoint, $p_{v}^{t2}$, must be present in the first timepoint all-lesion label, $y_{a}^{t1}$, but can not be present in the second timepoint all-lesion label, $y_{a}^{t2}$. We formulate these longitudinal constraints using a mean square error loss,
\begin{equation}
\label{eq:long}
\begin{aligned}
\mathcal{L}_{Long}=\frac{1}{N} \sum_{i=1}^{N} \|y^{t1}_{a_{i}}\odot p^{t2}_{n_{i}} - \mathbf{0}\|^2 + 
\frac{1}{N} \sum_{i=1}^{N} \|y^{t2}_{a_{i}}\odot p^{t2}_{n_{i}} - \mathbf{1}\|^2 + \\ 
\frac{1}{N} \sum_{i=1}^{N} \|y^{t1}_{a_{i}}\odot p^{t2}_{v_{i}} - \mathbf{1}\|^2 + 
\frac{1}{N} \sum_{i=1}^{N} \|y^{t2}_{a_{i}}\odot p^{t2}_{v_{i}} - \mathbf{0}\|^2,
\end{aligned}
\end{equation}
where $N$ is the total number of training images, $i$ denotes the image index, $\odot$ denotes voxel-wise multiplication, $\mathbf{0}$ denotes a zero matrix the size of the input image, $\mathbf{1}$ denotes an all-ones matrix, and $\|\cdot\|^2$ denotes the L2-norm for a flattened image. Differing from~\cite{Wu2023}, we impose longitudinal constraints on new and vanishing-lesion predictions rather than on all-lesion predictions. Where datasets are cross-sectional, longitudinal constraints are not imposed.

\subsubsection{Volumetric constraints}
Brain lesion volume changes over time. To account for the relationship of lesion volumes between two timepoints, we construct a volumetric constraint loss, which penalises high differences in lesion volume. We allow the lesion volume to increase or decrease~\cite{Sethi2017,Genovese2019, Filippi1995} within certain percentages, $\alpha_{high}$ and $\alpha_{low}$, and only apply the penalty if any change is beyond this. The volumetric constraint loss is
\begin{equation}
\label{eq:volumetric}
\mathcal{L}_{Vol} = \begin{cases*}
  \frac{1}{N} \sum_{i=1}^{N}(V_{a_{i}}^{t2} - \alpha_{high} \cdot V_{a_{i}}^{t1})^2 & if $ V_{a_{i}}^{t2} \geq \alpha_{high} \cdot V_{a_{i}}^{t1}$\\
  \frac{1}{N} \sum_{i=1}^{N}(V_{a_{i}}^{t2} - \alpha_{low} \cdot V_{a_{i}}^{t1})^2   & if $V_{a_{i}}^{t2}  \leq \alpha_{low} \cdot V_{a_{i}}^{t1} $\\
  0   & otherwise, 
\end{cases*}
\end{equation}
where $V_{a_{i}}^{t1}$ and $V_{a_{i}}^{t2}$ denote the total lesion volume at timepoint 1 and timepoint 2 for the $i$-th image, respectively. To our knowledge, there is no study on the proportion by how much lesion volume may \textit{decrease} with respect to time. We empirically set $\alpha_{low}$ to 0.8, after analysing the largest lesion volume decrease in a longitudinal MS dataset~\cite{Carass2015}. Similarly, we set $\alpha_{high}$ to 1.2, which conforms with medical literature~\cite{Molyneux1998}. $\alpha_{high}$ and $\alpha_{low}$ are set for annual percentage change. When a cross-sectional dataset is used, this penalty does not take effect.

\subsubsection{Spatial constraint}
MS primarily affects the white region of the brain, with lesions appearing hyperintense on FLAIR images~\cite{VanLeemput2001, Strumia2016, Lassmann2018}. We incorporate this prior in two ways: using the white matter mask, $x_{wm}^{t2}$, as an input channel to leverage the anatomical information; and also by constructing a spatial relation loss function, formulated as the mean square error between the predicted lesions, $p$, outside of the white matter, $x_{wm}^{t2}$, and a zero matrix the size of the image,
\begin{equation}
\label{eq:spatial}
\mathcal{L}_{Spat}=\frac{1}{N} \sum_{i=1}^{N} \|p\odot (\mathbf{1}-x_{wm_{i}}^{t2}) - \mathbf{0}\|^2.
\end{equation}

We employ a curriculum learning strategy~\cite{Bengio2009} and introduce the constraint losses after certain number of epochs. The total training loss function is
\begin{equation}
\label{eq:total}
\mathcal{L} = \begin{cases*}
  \mathcal{L}_{DSC} & if $n < \dfrac{n_{epoch}}{2}$,\\
  \mathcal{L}_{DSC} + \lambda_{L} \cdot \mathcal{L}_{Long} + \lambda_{V} \cdot \mathcal{L}_{Vol} + \lambda_{S} \cdot\mathcal{L}_{Spat}              & if $n \geq \dfrac{n_{epoch}}{2}$,
\end{cases*}
\end{equation}
where $\mathcal{L}_{DSC}$ denotes the Dice loss, $n$ is the epoch, $n_{epoch}$ denotes the total number of epochs for training, and $\lambda_{L}, \lambda_{V}$ and $\lambda_{S}$ are weighting parameters.
 
\section{Experiments}
\subsection{Data}
We curate five MS brain lesion datasets with data format or annotation heterogeneities. The first three are public datasets, including the ISBI 2015 MS lesion dataset (MS2015) \cite{Carass2015}, MICCAI 2016 MS lesion dataset (MS2016) \cite{Commowick2018}, and MICCAI 2021 MS new-lesion segmentation challenge dataset (MSSEG-2)~\cite{Commowick2021}. MS2015 and MS2016 contain all-lesion annotations. MSSEG-2 only contains new-lesion annotations at the second timepoint. A fourth dataset (MSSEG-2\Plus) is constructed based on MSSEG-2, which includes all-lesion annotations at the first timepoint, conducted by an experienced expert in-house using ITK-SNAP \cite{ITKSNAP}, along with new-lesion annotations at the second timepoint. Due to the lack of public datasets for vanishing lesions, a fifth dataset, (VAN), is constructed to simulate vanishing lesions by inverting the timepoints in MSSEG-2. Thus new lesions at the second timepoint become the lesions which will vanish from the first timepoint. These five datasets contain lesion images acquired from various MR scanners, patient cohorts, and with different annotation protocols. Further details of the datasets, including the training-test splits, are provided in Table~\ref{table:dataset_summary}. For all datasets, FLAIR images are used for lesion segmentation, which are resampled to $1\times1\times1$ mm$^3$ voxel spacing, followed by brain extraction using SynthSeg~\cite{Billot2023} and rigid registration into the MNI template space~\cite{Fonov2009}. 

\begin{table}[h!]
\setlength\tabcolsep{2pt}
\caption{Summary of the five datasets used for model training and evaluation, which contains different data formats (Image availability) and annotation styles (Label availability). $Y_{a}^{t1}$: first timepoint all lesion labels, $Y_{a}^{t2}$: second timepoint all-lesion labels, $Y_{n}^{t2}$: second timepoint new-lesion labels, $Y_{v}^{t2}$: second timepoint vanishing-lesion labels. $^{\dagger}$: Publicly available test set.}
\centering
\begin{tabular}[b]{p{16mm}C{10mm}C{10mm}C{15mm}C{15mm}C{10mm}C{10mm}C{10mm}C{10mm}}
\toprule
\multicolumn{3}{c}{Dataset} &\multicolumn{2}{c}{Image availability} & \multicolumn{4}{c}{Label availability} \\
\cmidrule(r){1-3}\cmidrule(lr){4-5}\cmidrule(l){6-9}
Name & Train\newline images & Test\newline images & Cross-\newline sectional & Longitudinal & $Y_{a}^{t1}$ & $Y_{a}^{t2}$ & $Y_{n}^{t2}$ & $Y_{v}^{t2}$ \\
\hline
MS2015 & 13 & 8 & - &\checkmark &\checkmark &\checkmark & - & -  \\
MS2016 & 11 & 4 & \checkmark & - & \checkmark & - & - & -  \\
MSSEG-2 & 40 & 60$^{\dagger}$  & - &\checkmark & - & - & \checkmark & - \\
MSSEG-2\Plus  & 40 & 60 & -  & \checkmark& \checkmark & - & \checkmark & - \\
VAN  & 40 & 60 & - &\checkmark& - & - & - & \checkmark \\
\hline
Total & 144 & 192 &&&&&&\\
\cmidrule[.75pt](r){1-3}
\label{table:dataset_summary}
\end{tabular}
\end{table}
\vspace{-1.2cm}
\subsection{Implementation details}
SegHeD is built on a 3D V-Net~\cite{Milletari2016} with four heads at the last layer for four lesion segmentation tasks. We utilise the V-Net architecture with five downsampling layers and an image patch size of $96\times96\times96$. The method is developed using PyTorch and NVIDIA Tesla T4 GPUs. We use the Adam optimizer with an initial learning rate of 0.001 and train for 20,000 epochs. $\lambda_{L}$, $\lambda_{V}$, and $\lambda_{S}$ are empirically set to 5, 1, and 1, respectively. Data augmentations include flipping, rotating, elastic deformation, additive and multiplicative brightness alterations, and additive Gaussian noise. Five-fold cross validation is conducted over a total of 144 training images, which results in an ensemble of five models. We report the performance of the ensemble on the test set. 

\subsection{Results}
\begin{figure}[h!] \centering
\includegraphics[width=.96\textwidth]{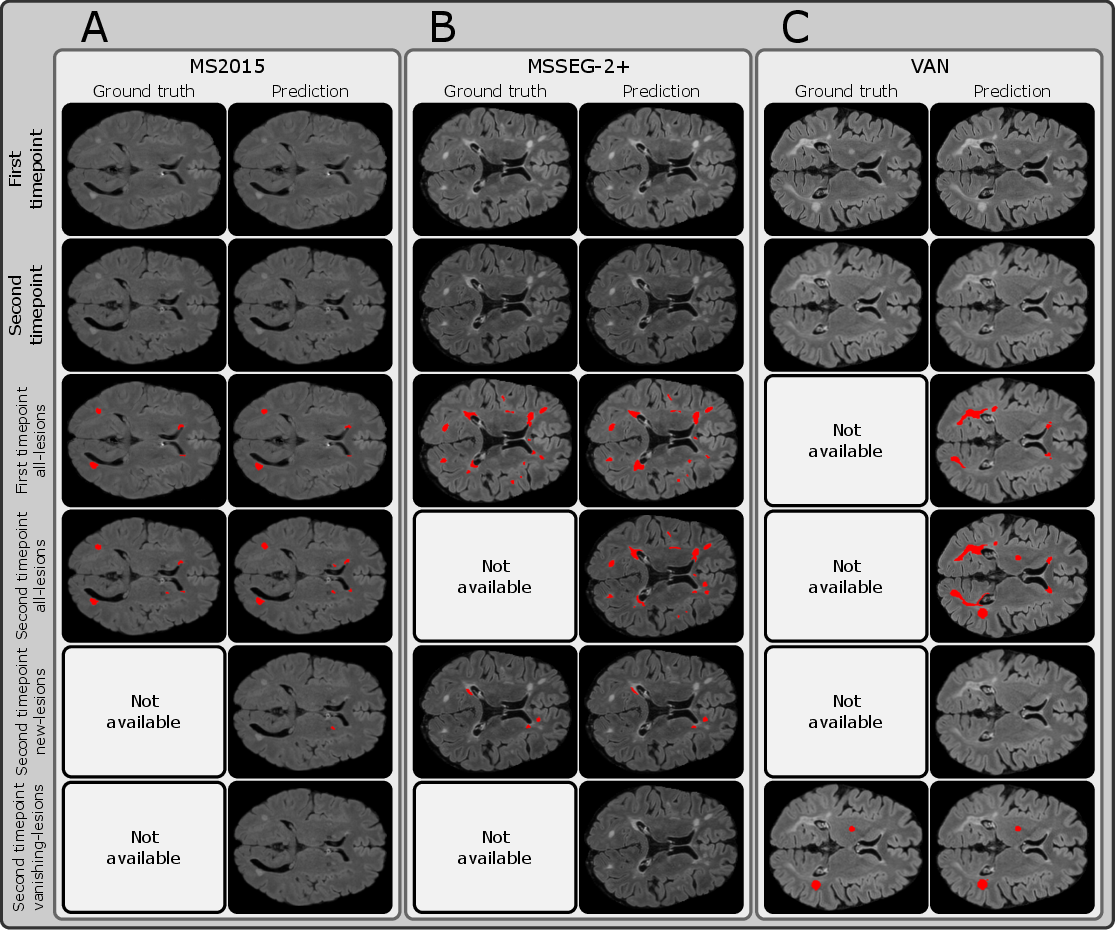}
\vspace{-.3cm}
\caption{SegHeD is capable of simultaneous multi-task segmentation (Rows 3 to 6). Some tasks do not show new/vanishing-lesions predictions as they are not present at the given slice. "Not available" denotes no ground truth annotation for comparison. \textbf{A:} Dataset where all-lesion labels are available for first and second timepoints. \textbf{B:}~Dataset where first timepoint all-lesion label and second timepoint new-lesion label are available. \textbf{C:} Dataset where second timepoint vanishing-lesion label is available.} 
\label{fig:segmentations_supp}
\end{figure}

SegHeD is a unified model that can perform all-lesion, new-lesion, and vanishing-lesion segmentation tasks. We evaluate its multi-task performance and compare it against state-of-the-art (SOTA) task-specific segmentation methods, nnU-net~\cite{Isensee2021}, nnFormer~\cite{Zhou2023}, UNETR~\cite{Hatamizadeh2022}, a recent heterogeneous data learning method, CoactSeg~\cite{Wu2023}, and specifically designed new-lesion segmentation methods \cite{zhang2021msseg, Basaran2022}, shown in Table~\ref{table:segresults}. Task-specific SOTA methods are trained twice, once for all-lesion segmentation task (using MS2015 and MS2016) and once for new-lesion segmentation task (using MSSEG-2). It is not possible to include the MSSEG-2\Plus dataset into SOTA training, as these methods do not allow for the heterogeneous annotations which MSSEG-2$\Plus$ possess. CoactSeg~\cite{Wu2023} is only trained once. For new lesion segmentation on MSSEG-2, we report Dice scores of MedICL~\cite{zhang2021msseg}, Basaran~\cite{Basaran2022}, and the average score of 4 human experts, \textit{Avg. of Experts}, officially released by the challenge organisers~\cite{Commowick2021}, and include the lesion-wise $F1$ scores in Table~\ref{table:segresults_f1}, in accordance with the MSSEG-2 challenge. Exemplar segmentations are compared in Figure~\ref{fig:segmentations}, and further provided in Figure~\ref{fig:segmentations_supp}.

\begin{table*}[h!] \centering
\setlength\tabcolsep{2pt}
\caption{Mean and standard deviations of lesion segmentation Dice scores ($\%$). Best results are in bold. N/A: output not available for the given method. $\dagger$: Methods where two models need to be trained, one for all-lesion and one for new-lesion segmentation. $\ddagger$: Not trained on MSSEG-2\Plus. Asterisks indicate statistical significance  ($^{*}$:~p$\leq$~0.05, $^{**}$:~p~$\leq$ 0.01, $^{***}$:~p~$\leq$ 0.005) when using a paired Student's \textit{t}-test comparing SegHeD's performance to benchmarked methods.}
\label{table:segresults}
\centering
\scalebox{1}{
\begin{tabular*}{\textwidth}{p{14mm} lccccc} \toprule
             & & \multicolumn{5}{c}{Segmentation task (lesion type)}\\
             \cmidrule(lr){3-7} 
             &  & \multicolumn{2}{c}{All} & \multicolumn{1}{c}{New} &\multicolumn{1}{c}{All\&New}  & \multicolumn{1}{c}{Vanishing} \\ 
            \cmidrule(lr){3-4} \cmidrule(lr){5-5} \cmidrule(lr){6-6} \cmidrule(lr){7-7} 
            Type & Method & \multicolumn{1}{c}{MS2015} & \multicolumn{1}{c}{MS2016} &\multicolumn{1}{c}{MSSEG-2}  & \multicolumn{1}{c}{MSSEG-2\Plus}  & \multicolumn{1}{c}{VAN} \\ 
     \cmidrule(lr){1-2} \cmidrule(lr){3-3} \cmidrule(lr){4-4} \cmidrule(lr){5-5} \cmidrule(lr){6-6} \cmidrule(lr){7-7} 
        \multirow{3}{*}{\shortstack[l]{New-\\ lesion\\methods}} & MedICL \cite{zhang2021msseg} & N/A & N/A & $50.67_{29.38}^{}$ & N/A & N/A \\ 
        &Basaran et al. \cite{Basaran2022}  & N/A & N/A & $51.06_{28.92}^{}$ & N/A & N/A \\ 
        &Avg. of Experts~\cite{Commowick2021}  & N/A & N/A & $\mathbf{55.52_{34.43}}$ & N/A  & N/A\\\midrule
        \multirow{3}{*}{\shortstack[l]{Task-\\ specific\\ SOTA}}&nnU-Net\cite{Isensee2021} $\dagger$ & $73.01_{4.91}^{***}$ & $74.87_{7.54}^{***}$ & \multicolumn{1}{|c}{$48.89_{31.20}^{}$} & N/A & N/A \\ 
        &nnFormer\cite{Zhou2023} $\dagger$ & $72.56_{7.15}^{***}$ & $74.12_{8.52}^{***}$ & \multicolumn{1}{|c}{$47.01_{33.39}^{*}$} & N/A & N/A \\
        &UNETR\cite{Hatamizadeh2022} $\dagger$ & $72.79_{6.13}^{***}$ & $73.78_{7.98}^{***}$ & \multicolumn{1}{|c}{$45.51_{30.84}^{**}$} & N/A & N/A \\ \midrule
        \multirow{2}{*}{\shortstack[l]{Hetero.\\methods}}&CoactSeg\cite{Wu2023} $\ddagger$  & $71.28_{8.24}^{***}$ & $71.31_{9.15}^{***}$ & $47.35_{38.12}^{}$ & $58.54_{18.54}^{***}$ & N/A \\
        &\textbf{SegHeD} & $\mathbf{78.10_{6.96}}$ & $\mathbf{84.73_{7.12}}$ &$48.64_{33.81}$ & $\mathbf{65.51_{19.67}}$ & $35.23_{20.62}$\\ \bottomrule
\end{tabular*}}
\end{table*}

\begin{table*}[h!] \centering
\caption{Mean and standard deviations of lesion detection $F1$ scores ($\%$), in accordance with the MSSEG-2 challenge~[7]. Best results are in bold. $\dagger$: Methods where two models are trained, one for all lesion and one for new lesion segmentation. $\ddagger$: Not trained on MSSEG\Plus. Asterisks indicate statistical significance  ($^{*}$:~p$\leq$~0.05, $^{**}$:~p~$\leq$ 0.01, $^{***}$:~p~$\leq$ 0.005) when using a paired Student's \textit{t}-test comparing SegHeD to competing methods.}
\label{table:segresults_f1}
\begin{tabular*}{\textwidth}{p{15mm}p{26.5mm}ccccc} \toprule
             \multicolumn{2}{c}{Methods}  & \multicolumn{2}{c}{All} & \multicolumn{1}{c}{New} &\multicolumn{1}{c}{All+New}  & \multicolumn{1}{c}{Vanishing} \\ 
            \cmidrule(r){1-2} \cmidrule(lr){3-4} \cmidrule(lr){5-5} \cmidrule(lr){6-6} \cmidrule(l){7-7} 
        \multirow{3}{*}{\shortstack[l]{New-\\ lesion\\methods}} & MedICL~\cite{zhang2021msseg} & N/A & N/A & $49.98_{32.78}^{*}$ & N/A & N/A \\ 
        &Basaran et al.~\cite{Basaran2022}  & N/A & N/A & $55.25_{34.81}^{}$ & N/A & N/A \\ 
        &Avg. of Expert~\cite{Commowick2021}  & N/A & N/A & $\mathbf{61.62_{37.11}}$ & N/A & N/A\\\midrule
        \multirow{3}{*}{\shortstack[l]{Task-\\ specific\\ SOTA}} & nnU-Net~\cite{Isensee2021} $\dagger$ & $75.81_{3.36}^{***}$ & $69.46_{11.44}^{***}$ & \multicolumn{1}{|c}{$54.15_{33.97}^{}$} & N/A & N/A \\
        &nnFormer~\cite{Zhou2023} $\dagger$ & $73.59_{5.04}^{***}$ & $68.13_{15.51}^{***}$ & \multicolumn{1}{|c}{$48.12_{33.53}^{}$} & N/A & N/A \\
        &UNETR~\cite{Hatamizadeh2022} $\dagger$ & $73.00_{4.92}^{***}$ & $70.12_{14.18}^{***}$ & \multicolumn{1}{|c}{$45.93_{39.24}^{**}$} & N/A & N/A \\
        \midrule
        \multirow{2}{*}{\shortstack[l]{Hetero.\\methods}}&CoactSeg~\cite{Wu2023} $\ddagger$  & $74.51_{4.77}^{***}$ & $69.23_{14.26}^{***}$ & $47.23_{36.45}^{*}$ & $54.31_{27.91}^{***}$ & N/A \\
        &\textbf{SegHeD} & $\mathbf{79.40_{3.27}}$ & $\mathbf{76.14_{10.97}}$ & $53.27_{34.70}$ & $\mathbf{59.20_{28.10}}$ & $39.48_{28.41}$\\ \bottomrule    
\end{tabular*}
\end{table*}

\subsubsection{All-lesion segmentation} 
\begin{figure}[h!] \centering
\includegraphics[width=.95\textwidth]{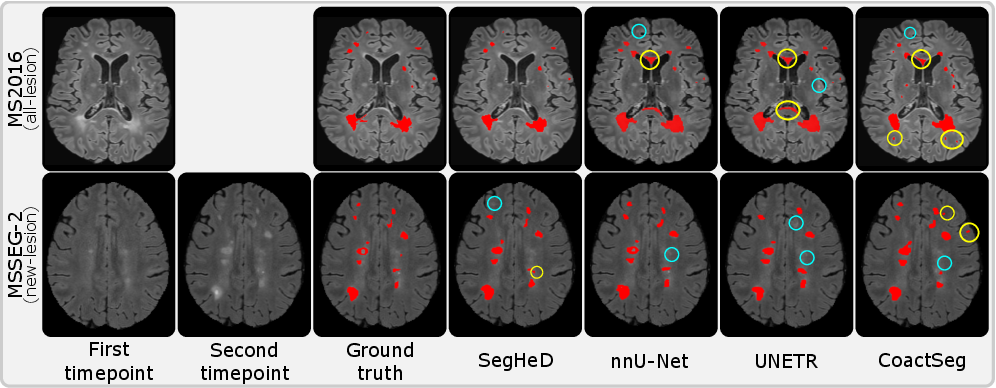}
\vspace{-.3cm}
\caption{Qualitative comparison of all-lesion (top row) and new-lesion (bottom row) segmentation performance. Yellow regions denote false positive segmentations, whereas cyan regions denote false negative segmentations.} 
\label{fig:segmentations}
\end{figure}

Table~\ref{table:segresults} shows that SegHeD significantly outperforms both task-specific SOTA methods and heterogeneous learning method CoactSeg in all-lesion segmentation task. For example, on MS2016 test set, SegHeD improves the Dice score by over 10\%. This can be attributed to several factors. SegHeD allows heterogeneous data input and thus can include more images for model training (Table~\ref{table:ablation_datasets}). The domain knowledge encoded via anatomical constraints also helps reduce false positives, as shown in Figure~\ref{fig:segmentations}. 

\subsubsection{New-lesion segmentation}
SegHeD performs competitively against task-specific SOTA methods and CoactSeg in new-lesion segmentation, achieving similar or slightly higher Dice scores. Although it slightly underperforms MedICL~\cite{zhang2021msseg} and Basaran~\cite{Basaran2022}, two top methods in MSSEG-2 challenge specifically designed for this dataset, SegHeD is capable of performing multiple tasks with a single model.

\subsubsection{Vanishing-lesion segmentation}
There are no existing methods dedicated for this objective. In this new yet challenging task, we achieve a Dice score of 35.23\%. The difficulty of this task is the joint modelling of vanishing lesions with all and new lesions. While task-specific methods need to learn either hyperintense region features for new and all-lesion segmentation, \emph{or} hypointense region features for vanishing lesion segmentation, SegHeD learns both features with respect to previous timepoints and surrounding tissues. We hope the results here provide useful insights for future dataset curation efforts and benchmarking work.

\newpage
\subsubsection{Ablation studies}
We perform two ablation studies for the different loss terms and the white matter mask input channel and the effect of incorporating heterogeneous data input. Table~\ref{table:ablation} shows that each component contributes to different tasks, while also increasing overall Dice scores when all incorporated. Notably, new and vanishing-lesion segmentation performance increases when $\mathcal{L}_{Long}$ is introduced, all-lesion segmentation performance increases when $\mathcal{L}_{Vol}$ is introduced, and an overall increase occurs when $\mathcal{L}_{Spat}$ and $x_{wm}^{t2}$ are presented. We also provide ablation results on SegHeD performance with varying amounts of heterogeneous data input in Table~\ref{table:ablation_datasets}. In particular, we observe a performance boost when datasets containing overlapping tasks are included. For instance, adding the MSSEG-2+ dataset, which specifically encompasses both the all-lesion and new-lesion segmentation tasks, leads to improvements in both all-lesion and new-lesion segmentation.
\vspace{-.5cm}
\begin{table}
\setlength\tabcolsep{4pt}
\caption{Ablation study of proposed losses and white matter mask input channel. Asterisks indicate statistical significance  ($^{*}$:~p$\leq$~0.05, $^{**}$:~p~$\leq$ 0.01, $^{***}$:~p~$\leq$ 0.005) when using a paired Student's \textit{t}-test comparing SegHeD's performance to ablated methods.}
\centering
\begin{tabular}[b]{ccccccccc}
\toprule
\multicolumn{4}{c}{Ablation Settings} &\multicolumn{5}{c}{Dice score (\%)} \\
\cmidrule(r){1-4} \cmidrule(l){5-9}
$\mathcal{L}_{Long}$ & $\mathcal{L}_{Vol}$ & $\mathcal{L}_{Spat}$ & $x_{wm}^{t2}$ & MS2015 & MS2016 & MSSEG-2 & MSSEG-2+ & VAN \\
\cmidrule(lr){1-1}  \cmidrule(lr){2-2} \cmidrule(lr){3-3} \cmidrule(lr){4-4} \cmidrule(lr){5-5} \cmidrule(lr){6-6} \cmidrule(lr){7-7} \cmidrule(lr){8-8} \cmidrule(lr){9-9} 
- & - & - & - & $73.19_{7.20}^{***}$ & $76.93_{9.01}^{***}$ & $45.28_{38.54}^{***}$ & $60.25_{20.04}^{***}$ & $30.97_{27.34}^{***}$\\
- & - & - & \checkmark & $73.59_{8.03}^{***}$ & $77.40_{9.05}^{***}$ & $46.15_{37.06}^{***}$ & $61.80_{20.90}^{***}$ & $31.50_{26.60}^{***}$\\
\checkmark & -  & - & - & $73.40_{6.88}^{***}$ & $77.38_{8.99}^{***}$ & $48.55_{34.80}$ & $64.86_{21.12}^{*}$ & $33.88_{28.97}^{**}$\\
\checkmark & \checkmark & - & - & $75.91_{7.59}^{**}$ & $81.99_{8.03}^{**}$ & $48.49_{35.00}$ & $65.01_{20.21}^{*}$ & $34.52_{30.63}^{*}$\\
\checkmark & \checkmark & \checkmark & - & $77.93_{6.72}$ & $82.71_{7.42}^{**}$& $48.56_{34.04}$ & $65.20_{19.79}^{*}$ & $34.90_{28.57}^{*}$\\
\checkmark & \checkmark & \checkmark & \checkmark & $78.10_{6.96}$ & $84.73_{7.12}$ &$48.64_{33.81}$ & $65.51_{19.07}$ & $35.23_{20.62}$\\
\bottomrule
\label{table:ablation}
\end{tabular}
\end{table}
\vspace{-1cm}
\begin{table*}[h!]
\setlength\tabcolsep{2pt}
\caption{Ablation study to show the improved performance when including additional heterogeneous datasets. All studies are implemented with longitudinal reasoning. N/A: output not available for the given dataset.}
\centering
\begin{tabular*}{\textwidth}{C{5mm}C{5mm}C{5mm}C{5mm}C{5mm}C{17mm}C{17mm}C{17mm}C{17mm}C{15mm}} 
\toprule
\multicolumn{5}{c}{Datasets included} &\multicolumn{5}{c}{Dice score (\%)} \\
\cmidrule(r){1-5} \cmidrule(l){6-10}
\multirow{3}{*}{\vspace{-.7cm}\rotatebox[origin=c]{90}{\footnotesize  MS2015}} & 
\multirow{3}{*}{\vspace{-.7cm}\rotatebox[origin=c]{90}{\footnotesize MS2016}} & 
\multirow{3}{*}{\vspace{-.5cm}\rotatebox[origin=c]{90}{\footnotesize MSSEG-2}} & 
\multirow{3}{*}{\vspace{-.35cm}\rotatebox[origin=c]{90}{\footnotesize MSSEG-2\Plus}} & 
\multirow{3}{*}{\vspace{-1.05cm}\rotatebox[origin=c]{90}{\footnotesize VAN}} & 
\multicolumn{5}{c}{Segmentation task (lesion type)} \\
\cmidrule(lr){6-10}
& & & & & \multicolumn{2}{c}{All} & \multicolumn{1}{c}{New} &\multicolumn{1}{c}{All\&New}  & \multicolumn{1}{c}{Vanishing} \\
\cmidrule(lr){6-7} \cmidrule(lr){8-8} \cmidrule(lr){9-9} \cmidrule(lr){10-10} 
& & & & & MS2015 & MS2016 & MSSEG-2 & MSSEG-2\Plus & VAN \\ 
\midrule
\checkmark & \checkmark & - & - & - & $74.97_{5.35}$ & $77.92_{7.40}$ & N/A & N/A & N/A \\
- & - & \checkmark & - & - & N/A & N/A & $49.00_{32.32}$ & N/A & N/A\\
\checkmark & \checkmark & \checkmark & - & - & $75.30_{6.51}$ & $78.11_{8.42}$ & $46.73_{34.17}$ & $61.16_{21.50}$ & N/A\\
\checkmark & \checkmark & \checkmark & \checkmark & - & $77.93_{7.39}$ & $84.76_{7.19}$ & $48.02_{35.27}$ & $65.88_{19.20}$ & N/A \\
\checkmark & \checkmark & \checkmark & \checkmark & \checkmark & $78.10_{6.96}$ & $84.73_{7.12}$ & $48.64_{33.81}$ & $65.51_{19.67}$ & $35.23_{20.62}$\\
\bottomrule
\label{table:ablation_datasets}
\end{tabular*}
\end{table*}

\subsubsection{Improved temporal consistency}
\begin{figure}[h!] \centering
\includegraphics[width=\textwidth]{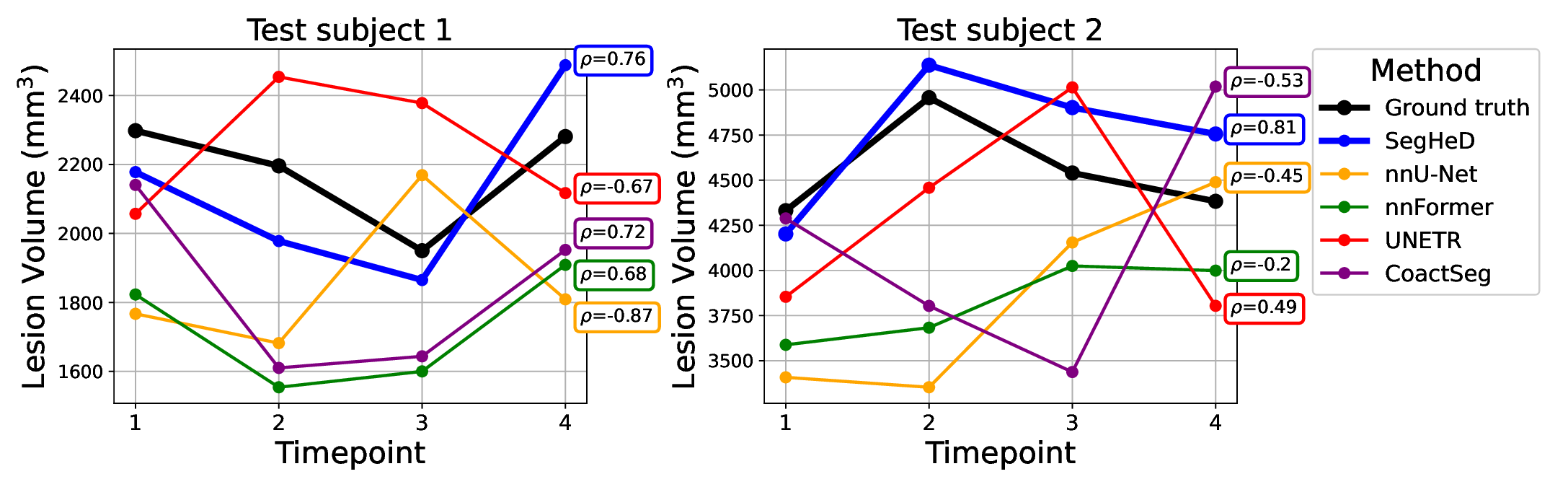}
\caption{Predicted lesion volumes across four timepoints for two test subjects. SegHeD (blue) predictions are temporally more consistent with the ground truth (black), compared to competing methods. The $\rho$ value for each method indicates its Pearson's correlation coefficient with the ground truth, the higher the better.} 
\label{fig:consistency}
\end{figure}

Fig.~\ref{fig:consistency} displays predicted lesion volumes across multiple timepoints for two test subjects from MS2015. SegHeD predictions (blue) show higher temporal consistency with the ground truth (black) due to the proposed constraints, obtaining the highest Pearson's correlation coefficient with the ground truth. Improved temporal consistency facilitates downstream analysis tasks, such as evaluating annual lesion atrophy or growth rate.

\section{Conclusion} 
We present SegHeD, a novel multi-task MS lesion segmentation method for learning from heterogeneous and longitudinal data. It is capable of segmenting all, new, and vanishing lesions for both cross-sectional and longitudinal datasets. Experiments on five MS lesion datasets show that SegHeD outperforms other competing segmentation methods for all-lesion segmentation, and performs competitively for new-lesion segmentation. Notably, its capability of leveraging heterogeneous data will greatly advance existing MS imaging studies and facilitate large-scale multi-site data analyses with diverse data formats and annotations.

\subsubsection{Acknowledgements} This work is supported by the UKRI CDT in AI for Healthcare \href{http://ai4health.io}{http://ai4health.io} (Grant No. EP/S023283/1). W. Bai is co-funded by EPSRC DeepGeM Grant (EP/W01842X/1) and NIHR Imperial Biomedical Research Centre (BRC). The views expressed are those of the authors and not necessarily those of the NIHR or the Department of Health and Social Care. 

%
%
\newpage
\bibliographystyle{splncs04}
\bibliography{Paper-0002}

\end{document}